\documentclass[letterpaper]{article} 
\usepackage{aaai2026}  
\usepackage{times}  
\usepackage{helvet}  
\usepackage{courier}  
\usepackage[hyphens]{url}  
\usepackage{graphicx} 
\usepackage{natbib}  
\usepackage{caption} 
\usepackage[table]{xcolor}

\definecolor{RoyalBlue}{HTML}{4169E1}      
\definecolor{CornflowerBlue}{HTML}{6495ED} 
\definecolor{SeaGreen}{HTML}{2E8B57}       
\definecolor{LightSeaGreen}{HTML}{20B2AA}  
\definecolor{Goldenrod}{HTML}{DAA520}      
\definecolor{Orchid}{HTML}{DA70D6}         
\definecolor{MediumOrchid}{HTML}{BA55D3}   
\definecolor{Plum}{HTML}{DDA0DD}           
\definecolor{Coral}{HTML}{FF7F50}          
\definecolor{Tomato}{HTML}{FF6347}         

\newcommand{\blue}[1]{\textcolor{RoyalBlue}{#1}}
\newcommand{\green}[1]{\textcolor{SeaGreen}{#1}}
\newcommand{\purple}[1]{\textcolor{MediumOrchid}{#1}}
\newcommand{\orange}[1]{\textcolor{Goldenrod}{#1}}

\usepackage{algorithm}
\usepackage{algorithmic}
\usepackage{amsmath}
\usepackage{amsfonts}

\usepackage{tcolorbox}
\tcbuselibrary{theorems,skins,breakable}

\newtcolorbox{keybox}[1][]{
    colback=CornflowerBlue!10,
    colframe=RoyalBlue!80,
    fonttitle=\bfseries,
    title={#1},
    boxrule=1.5pt,
    arc=3mm,
    left=5pt,
    right=5pt,
    top=5pt,
    bottom=5pt,
    breakable
}

\newtcolorbox{highlightbox}[1][]{
    colback=SeaGreen!8,
    colframe=SeaGreen!70,
    fonttitle=\bfseries,
    title={#1},
    boxrule=1.2pt,
    arc=2mm,
    left=5pt,
    right=5pt,
    top=4pt,
    bottom=4pt,
    breakable
}

\newcommand{\method}[1]{\textbf{#1}} 
\newcommand{\dataset}[1]{\texttt{#1}} 
\newcommand{\metric}[1]{\textit{#1}} 

\usepackage{newfloat}
\usepackage{listings}
\DeclareCaptionStyle{ruled}{labelfont=normalfont,labelsep=colon,strut=off} 
\floatstyle{ruled}
\newfloat{listing}{tb}{lst}{}
\floatname{listing}{Listing}

\title{Aligning Generative Music AI with Human Preferences: Methods and Challenges}
\author{
    Dorien Herremans, Abhinaba Roy \\ 
}
\affiliations{
    AMAAI Lab, Singapore University of Technology and Design\\

    8 Somapah Road\\
    487372 Singapore\\
    dorien\_herremans@sutd.edu.sg, abhinaba\_roy@sutd.edu.sg
}

\usepackage{bibentry}

\begin{document}

\maketitle

\begin{abstract}

Recent advances in generative AI for music have achieved remarkable fidelity and stylistic diversity, yet these systems often fail to align with nuanced human preferences due to the specific loss functions they use. This paper advocates for the systematic application of preference alignment techniques to music generation, addressing the fundamental gap between computational optimization and human musical appreciation. Drawing on recent breakthroughs including MusicRL's large-scale preference learning, multi-preference alignment frameworks like diffusion-based preference optimization in DiffRhythm+, and inference-time optimization techniques like Text2midi-InferAlign, we discuss how these techniques can address music's unique challenges: temporal coherence, harmonic consistency, and subjective quality assessment. We identify key research challenges including scalability to long-form compositions, reliability amongst others in preference modelling. Looking forward, we envision preference-aligned music generation enabling transformative applications in interactive composition tools and personalized music services. This work calls for sustained interdisciplinary research combining advances in machine learning, music-theory to create music AI systems that truly serve human creative and experiential needs.
\end{abstract}


\section{Introduction}

\begin{figure*}[t]
\centering
\includegraphics[width=0.93\textwidth]{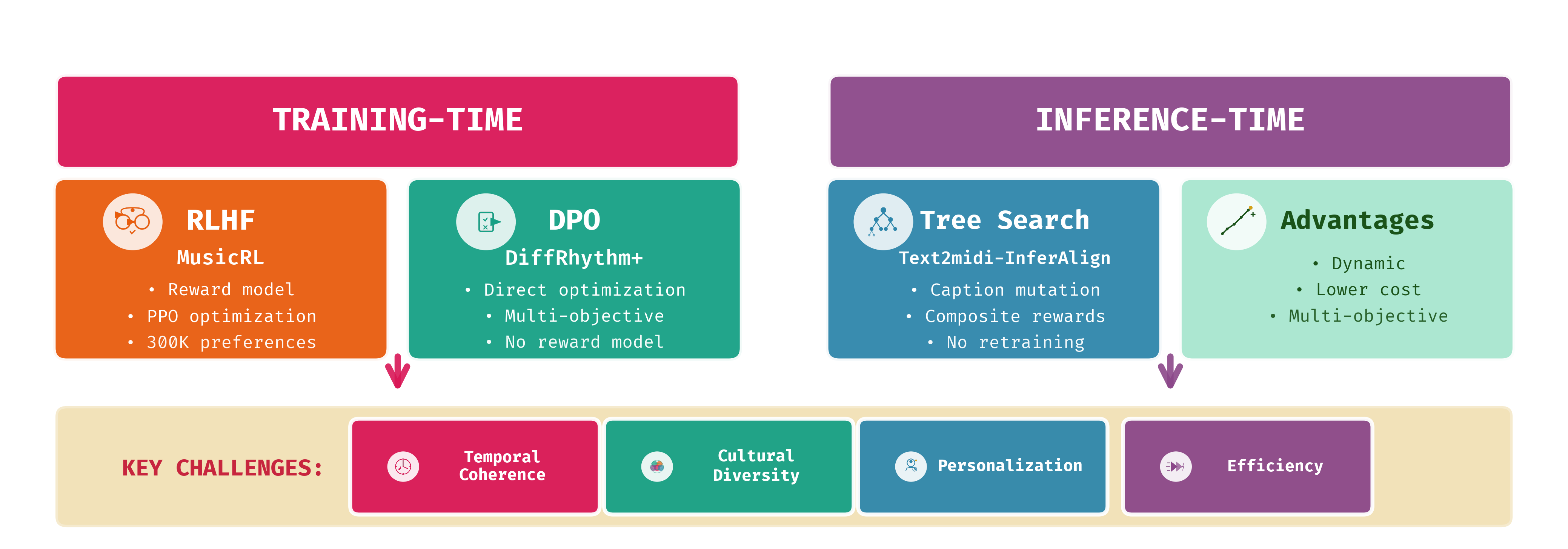}
\caption{Overview of selected preference alignment approaches for music generation. Training-time methods (left) optimize models during development, while inference-time methods (right) adapt generation without retraining. Both address the fundamental gap between likelihood-based training and human musical preferences.}
\label{fig:framework}
\end{figure*}




The emergence of large-scale generative models for music has transformed the landscape of artificial creativity, achieving unprecedented fidelity in audio synthesis and remarkable diversity in stylistic expression. Contemporary systems such as MusicLM \citep{agostinelli2023musiclm}, MusicGen \citep{copet2023musicgen}, Mustango \cite{melechovsky2024mustango} and Jukebox \citep{dhariwal2020jukebox} demonstrate capabilities that would have been unbelievable just years ago, generating high-quality musical compositions from textual descriptions with impressive technical proficiency. Yet despite these remarkable achievements, these systems face a fundamental and persistent challenge: aligning their outputs with the nuanced, subjective nature of human musical preferences.

Traditional supervised training approaches in music generation rely predominantly on likelihood-based objectives that optimize for training distributions. While these objectives successfully capture surface-level musical patterns and enable coherent generation, they fundamentally fail to capture the deeper qualities that make music aesthetically pleasing, emotionally resonant to human listeners \citep{juslin2008emotional}. This alignment gap becomes particularly pronounced in music generation, where subjective appreciation depends on complex, multi-layered factors including cultural context, emotional resonance, personal taste, listening environment, and the intricate interplay between musical elements across multiple temporal scales \cite{agres2016evaluation}.

The challenges inherent to musical preference alignment extend far beyond those encountered in other generative domains. Unlike text generation where semantic correctness and factual accuracy provide objective evaluation criteria \citep{blei2003latent}, or image generation where visual fidelity offers measurable benchmarks \citep{salimans2016improved, heusel2017gans}, musical quality encompasses a complex constellation of factors that resist straightforward quantification. These include temporal coherence across extended sequences that may span minutes or hours, harmonic consistency that must maintain theoretical soundness while supporting creative expression, rhythmic stability that provides structural foundation while allowing for expressive variation, timbral quality that affects emotional perception \citep{mcadams2013musical}, and subjective aesthetic judgment that varies dramatically across individuals and cultural contexts \citep{north2016music}.

The concept of quality of music itself defies universal definition, being inherently user-dependent and context-sensitive. A caption such as ``upbeat workout music" might legitimately map to diverse interpretations ranging from retro guitar solos to techno pop beats, from orchestral arrangements to electronic dance music, each carrying different emotional connotations and serving different functional purposes \citep{cideron2024musicrl}. This inherent ambiguity in musical interpretation challenges traditional supervised learning paradigms and calls for specialized approaches that can integrate continuous human feedback into post-deployment model refinement.

Furthermore, music exists within complex social and cultural ecosystems that shape preference formation in ways that purely technical metrics cannot capture. Musical preferences are influenced by factors including cultural background, age demographics, social identity, historical context, and personal experiences that create deep emotional associations with particular musical styles or elements \citep{north2016music}. These preferences evolve dynamically over time, both at individual and collective levels, requiring adaptive systems that can respond to changing aesthetic landscapes \citep{schedl2014music}.

Recent breakthroughs have demonstrated promising directions for addressing these multifaceted challenges through sophisticated preference alignment techniques. MusicRL \citep{cideron2024musicrl} is the first music generation system fine-tuned from human feedback at scale, collecting an unprecedented dataset of approximately 300,000 pairwise preferences to train models that significantly outperform baselines in comprehensive human evaluations. However, this preference dataset remains proprietary and is not publicly available, limiting reproducibility and broader adoption. This work demonstrates that large-scale preference learning can bridge the gap between computational optimization and human musical appreciation, while revealing the complex, multi-dimensional nature of musical preference that extends beyond simple quality metrics.

DiffRhythm+ \citep{chen2025diffrhythm+} extends preference optimization to diffusion-based architectures, enabling multi-preference alignment for controllable full-length song generation that can simultaneously satisfy multiple preference criteria. This work demonstrates how modern generative architectures can be adapted to incorporate preference signals at multiple levels of musical structure, from local harmonic progressions to global compositional form.

Text2midi-InferAlign \citep{roy2025text2midi} introduces inference-time optimization techniques that achieve substantial improvements in text-audio consistency without requiring expensive model retraining, opening new possibilities for dynamic preference adaptation. Through sophisticated tree search algorithm, this approach demonstrates that preference alignment can be achieved through clever inference-time computation, enabling real-time adaptation to user preferences and contextual requirements.

The field builds upon earlier foundational work in reinforcement learning for music generation. \citet{jaques122016generating} used RL-based music generation by fine-tuning RNN-based models using rewards incorporating music theory rules, while \citet{le2010towards} developed adaptive interactive systems using RL agents for real-time emotional feedback optimization. Recent systems have continued this trajectory: JAM \citep{liu2025jam} employs DPO, NotaGen \citep{wang2025notagen} introduces CLaMP-DPO for classical sheet music generation, and inference-time methods like DITTO \citep{novack2024ditto} and SMITIN \citep{koo2025smitin} demonstrate sophisticated optimization strategies during generation.

These advances collectively highlight three core technical approaches to preference alignment in music generation as shown in Figure~\ref{fig:framework}: (1) \method{large-scale preference learning} through reinforcement learning from human feedback (RLHF) that can capture complex preference patterns from extensive human evaluation data, (2) \method{inference-time optimization} that balances multiple objectives during generation without requiring model modification, and (3) \method{direct preference optimization} integrated into modern generative architectures that can learn preference-aligned representations during training. Each approach offers distinct advantages and faces unique challenges, creating a rich landscape of technical possibilities for future development.

We provide a comprehensive examination of how these methods address music's fundamental challenges while identifying key research directions for creating music AI systems that truly serve human creative needs. We argue that the future of music generation lies not in technical sophistication or computational scale, but in meaningful alignment with the rich complexity of human musical preference, cultural expression, and creative intention. 




\section{Background}

Preference alignment in generative models has emerged as a critical paradigm for creating systems that produce outputs aligned with human judgment rather than merely maximizing likelihood with respect to training distributions. This shift represents a fundamental reconceptualization of the generative modeling objective, moving from statistical fidelity to human-centered quality optimization \citep{christiano2017deep, bai2022constitutional}.

\subsection{Reinforcement Learning from Human Feedback}

Reinforcement Learning from Human Feedback (RLHF) \citep{christiano2017deep} pioneered the preference alignment paradigm by decomposing the alignment problem into two distinct phases. In the first phase, a \textbf{reward model} $r_\phi(x, y)$ is trained to predict human preferences using a dataset of preference comparisons $\mathcal{D} = \{(x_i, y_i^w, y_i^l)\}_{i=1}^N$, where $y_i^w$ represents the preferred output and $y_i^l$ the less preferred output for input $x_i$. The reward model is typically trained using the Bradley-Terry preference model \citep{bradley1952rank}:

$$p(y^w \succ y^l | x) = \sigma(r_\phi(x, y^w) - r_\phi(x, y^l))$$

where $\sigma$ denotes the sigmoid function and $\succ$ indicates preference ordering.

In the second phase, the generative policy $\pi_\theta$ is optimized to maximize expected rewards while maintaining similarity to a reference policy $\pi_{\text{ref}}$ through a regularization term:

$$\max_\theta \mathbb{E}_{x \sim \mathcal{D}, y \sim \pi_\theta(\cdot|x)}[r_\phi(x, y)] - \beta \mathbb{E}_{x \sim \mathcal{D}}[\text{KL}(\pi_\theta(\cdot|x) \| \pi_{\text{ref}}(\cdot|x))]$$

where $\beta$ controls the strength of the Kullback–Leibler(KL) regularization. This objective is typically optimized using policy gradient methods such as \method{Proximal Policy Optimization (PPO)} \citep{schulman2017proximal}.

While \method{RLHF} has demonstrated remarkable success across multiple domains, it suffers from several limitations including training instability, high computational requirements due to the need for multiple model training phases, and potential reward hacking where the policy learns to exploit weaknesses in the reward model rather than genuinely improving quality \citep{bai2022constitutional}.

\subsection{Direct Preference Optimization}

\method{Direct Preference Optimization (DPO)} \citep{rafailov2023direct} addresses the limitations of RLHF by \textbf{eliminating the explicit reward model} training phase. DPO leverages a key theoretical insight: under the optimal policy for the RLHF objective, a closed-form solution exists between the optimal policy, the reference policy, and the implicit reward function.

Specifically, if $\pi^*$ represents the optimal policy for the RLHF objective, then:

$$\pi^*(y|x) = \frac{1}{Z(x)} \pi_{\text{ref}}(y|x) \exp\left(\frac{1}{\beta} r(x, y)\right)$$

where $Z(x)$ is a partition function. Rearranging this relationship yields:

$$r(x, y) = \beta \log \frac{\pi^*(y|x)}{\pi_{\text{ref}}(y|x)} + \beta \log Z(x)$$

Since the partition function is independent of $y$, preference comparisons eliminate this term, allowing DPO to directly optimize the policy using the reparameterized objective:


\begin{align*}
\mathcal{L}_{\text{DPO}}(\pi_\theta) =\ & -\mathbb{E}_{(x,y^w,y^l) \sim \mathcal{D}}\Bigg[\log \sigma\Bigg(\beta \log \frac{\pi_\theta(y^w|x)}{\pi_{\text{ref}}(y^w|x)} \\
&\qquad - \beta \log \frac{\pi_\theta(y^l|x)}{\pi_{\text{ref}}(y^l|x)}\Bigg)\Bigg]
\end{align*}

This formulation enables stable optimization while maintaining the expressiveness of reward-based methods, offering significant computational advantages over traditional RLHF approaches.

\subsection{Inference-Time Alignment Paradigms}

Inference-time alignment techniques offer complementary advantages by enabling \textbf{preference optimization without model retraining}. These methods modify the generation process itself to satisfy preference constraints through techniques such as:

\textbf{\method{Contrastive Decoding:}} Amplifying probability mass from preferred directions while suppressing undesired outputs through contrastive search procedures that balance exploration and exploitation during generation \citep{li2023inference}.

\textbf{\method{Preference-Conditioned Sampling:}} Modifying sampling distributions to favor outputs that satisfy specified preference criteria, often through energy-based reweighting or guided generation techniques \citep{bai2022constitutional}.

\textbf{\method{Control Vector Steering:}} Manipulating intermediate representations within generative models to guide outputs toward desired characteristics without explicit retraining \citep{subramani2022extracting}.

In music generation, this flexibility proves essential for balancing competing preferences such as text adherence, audio quality, stylistic consistency, and temporal coherence. The ability to dynamically adjust these trade-offs during inference enables responsive adaptation to user preferences and contextual requirements.

\section{Musical Preference Complexity}

In the music domain, traditional evaluation metrics such as \metric{Fréchet Audio Distance (FAD)}, \metric{Inception Score (IS)}, or \metric{CLAP-based text-audio similarity} fail to capture the subjective nature of musical appreciation. Musical preferences emerge from complex interactions between multiple dimensions:

\textbf{Temporal Structure:} Music unfolds over time with hierarchical patterns spanning multiple scales from local rhythmic patterns (milliseconds to seconds) to phrase-level structure (seconds to minutes) to global compositional form (minutes to hours) \cite{lerdahl1996generative}. Preference alignment must account for coherence across all these temporal scales simultaneously.

\textbf{Harmonic Content:} Western tonal music relies on sophisticated harmonic relationships that create expectations and resolutions. Preferences often depend on whether these harmonic progressions feel satisfying, surprising, or emotionally appropriate within their musical context \cite{meyer1954emotion}.

\textbf{Cultural Context:} Musical preferences are deeply embedded within cultural frameworks that shape aesthetic expectations. What constitutes "good" music varies dramatically across musical traditions, historical periods, and social communities \cite{liu2018relation}.

\textbf{Functional Considerations:} Music serves diverse functional roles—from background ambiance to focused listening, from dance accompaniment to game music and emotional expression—each requiring different optimization criteria \cite{herremans2017functional}.

These factors necessitate specialized alignment approaches that can handle the multi-dimensional nature of musical preference while respecting the inherent subjectivity and cultural specificity of musical aesthetics.

\section{Preference Optimization in Music}

Application of preference optimization to music generation presents unique opportunities and challenges that distinguish it from text-based preference alignment. We discuss three prominent approaches here. 

\subsection{Large-Scale Preference Learning with MusicRL}

MusicRL \citep{cideron2024musicrl} pioneered \textbf{large-scale preference learning} in music generation by fine-tuning a pretrained autoregressive MusicLM model using both expert-designed reward functions and massive human preference datasets. The system represents a groundbreaking implementation that demonstrates how preference alignment can bridge the gap between computational optimization and human musical appreciation at unprecedented scale.

The MusicRL framework implements two complementary optimization strategies that address different aspects of musical preference alignment. MusicRL-R employs explicit reward functions designed in collaboration with expert raters, focusing on sequence-level rewards related to text adherence and audio quality. These reward functions capture measurable aspects of musical quality including semantic alignment between textual descriptions and generated audio, perceptual audio quality measures that assess technical fidelity and artifacts, and musical structure coherence, evaluating temporal consistency and harmonic progression.

MusicRL-U incorporates human feedback at scale through Reinforcement Learning from Human Feedback (RLHF). The system collected a private dataset containing pairwise preferences from real users, creating the largest known dataset of human preferences for music generation. This massive preference collection enables training a sophisticated preference model that captures subtle aspects of subjectivity.

The preference data collection process involved presenting users with pairs of generated musical pieces and asking them to indicate preferences based on overall quality, musical coherence, and text adherence. This process revealed important insights about musical preference: human evaluators consistently demonstrate preferences that extend beyond simple audio quality or text matching, indicating aesthetic judgments that traditional metrics fail to capture.

Experimental results demonstrate that both MusicRL-R and MusicRL-U significantly outperform baseline MusicLM in human evaluations across diverse musical styles. The combined MusicRL-RU approach achieves the strongest performance. Crucially, ablation studies reveal that text adherence and audio quality account for only a portion of human preferences, underscoring the prevalence of subjectivity in musical appreciation and highlighting the inadequacy of current evaluation metrics.

\subsection{Multi-Preference Diffusion Optimization with DiffRhythm+}

DiffRhythm+ \citep{chen2025diffrhythm+} extends preference optimization to \textbf{diffusion-based architectures} for controllable full-length song generation, demonstrating how Direct Preference Optimization (DPO) can be effectively integrated into modern generative frameworks. The system employs multi-modal style conditioning through MuLan embeddings, enabling precise control over musical attributes while maintaining preference alignment.

The framework integrates DPO directly into the diffusion training process using a modified DPO objective adapted for continuous latent spaces. Unlike discrete sequence models, diffusion models operate in continuous spaces, requiring careful adaptation of preference optimization techniques.

DiffRhythm+ incorporates metrics from multiple evaluation frameworks including \metric{SongEval} \cite{yao2025songeval} and \metric{Audiobox-aesthetic} \cite{tjandra2025meta}, enabling optimization for diverse preference criteria simultaneously. SongEval provides a more comprehensive assessment that includes structural coherence and memorability, evaluating how well generated music maintains logical harmonic progressions and consistent rhythmic patterns across extended compositions. Audiobox-aesthetic metrics focus on perceptual quality and aesthetic appeal, incorporating learned models that predict human aesthetic judgments.

The integration of DPO into diffusion architectures presents unique advantages for music generation. Unlike autoregressive models that generate sequentially, diffusion architectures can optimize global structure and long-range dependencies simultaneously, proving particularly valuable for music where harmonic progressions and rhythmic patterns must maintain coherence across extended temporal spans. Experimental validation demonstrates substantial improvements over baseline diffusion models, with particularly strong performance in long-form musical coherence and aesthetic appeal \citep{chen2025diffrhythm+}.

Both MusicRL and DiffRhythm+ demonstrate the critical importance of multi-objective optimization in musical preference alignment. Music generation must simultaneously satisfy competing constraints across multiple dimensions that often conflict with each other, requiring sophisticated balancing mechanisms.


The success of these methods shows that DPO can effectively balance generation objectives while maintaining diversity, essential for creative applications. However, the complexity of multi-objective optimization in music generation suggests opportunities for more sophisticated preference modeling techniques that can explicitly represent and balance competing criteria during optimization.

\section{Inference-Time Alignment Techniques}

Text2midi-InferAlign \citep{roy2025text2midi} demonstrates sophisticated inference-time preference optimization through a \method{tree search methodology} that balances multiple reward objectives at inference. The framework employs a composite reward function combining (weighted) text-audio consistency scores ($S_{\text{text}}(y_t, x)$) with harmonic consistency scores ($S_{\text{harmony}}$):

$$\text{Score}(y_t, x) = \alpha \cdot S_{\text{text}}(y_t, x) + \beta \cdot S_{\text{harmony}}(y_t) $$

The system incorporates caption mutation for exploration, generating semantically related variations of input captions to explore different musical interpretations while maintaining core meaning. The reward mechanism efficiently navigates the search space without exhaustive enumeration.

Experimental results demonstrate effectiveness: this approach achieves \textbf{\blue{29.4\% improvement}} in CLAP scores compared to baseline text2midi generation \cite{bhandari2025text2midi} without any model modification, preserving diversity while enhancing quality. However, the tree search process requires some computational overhead, creating trade-offs between quality improvements and inference latency that future work must address for real-time applications.
\begin{table*}[ht!]
\centering
\caption{Key Research Challenges in Music Preference Alignment}
\label{tab:challenges}
\begin{tabular}{|l|p{0.7\textwidth}|}
\hline
\rowcolor{RoyalBlue!15}
\textbf{Challenge} & \textbf{Key Issues} \\
\hline
\textbf{Scalability} & Long-form composition modeling, attention complexity, hierarchical structures \\
\hline
\rowcolor{SeaGreen!10}
\textbf{Multi-modal} & Video-music sync, cross-cultural integration, real-time adaptation \\
\hline
\textbf{Personalization} & Few-shot learning, individual aesthetics, cultural awareness \\
\hline
\rowcolor{Coral!10}
\textbf{Robustness} & Adversarial attacks, bias amplification, quality degradation \\
\hline
\textbf{Efficiency} & Inference overhead, energy costs, interactivity latency \\
\hline
\rowcolor{Orchid!10}
\textbf{Evaluation} & Preference learning, cross-domain transfer, evaluation paradox \\
\hline
\end{tabular}
\end{table*}

\section{Benchmarking and Evaluation}

Evaluating preference-aligned music generation requires metrics that capture both objective technical quality and subjective human judgment. Traditional metrics like \metric{Fréchet Audio Distance (FAD)} \citep{kilgour2018fr} and \metric{Inception Score (IS)} \citep{barratt2018note} provide baselines but fail to capture musical-specific qualities or aesthetic appeal.

Recent work introduces dedicated frameworks: MusicRL's pairwise preferences data demonstrates that automated metrics correlate only partially with human judgment. Text2midi-InferAlign employs \metric{CLAP scores} for text-audio consistency alongside harmonic consistency metrics. DiffRhythm+ incorporates \metric{SongEval} and \metric{Audiobox-aesthetic} frameworks for comprehensive quality assessment.

Key evaluation challenges include developing metrics for diverse musical traditions and tastes, creating standardized benchmarks for long-form musical coherence, and establishing protocols for personalized preference evaluation. Cross-cultural validity remains important as many current frameworks often reflect recent western musical content (rock, pop, electronic primarily), limiting global applicability.

\section{Technical Implementation Considerations}

Practical deployment of preference-aligned music generation requires addressing computational scalability, data management, and user interface design. Large-scale models like MusicLM require billions of parameters, while preference learning adds overhead through \method{RLHF} (or \method{DPO}) training and \method{inference-time optimization} creates additional computational costs at inference.

The architectural considerations for preference alignment systems extend beyond model parameters to encompass sophisticated data pipelines and training infrastructure. MusicRL's success with preference data required specialized data collection platforms capable of presenting musical pairs to human evaluators while maintaining consistent quality assessment protocols. The system architecture must also have quality control mechanisms to identify inconsistent evaluators, and sophisticated aggregation algorithms that account for individual biases and cultural backgrounds.
\begin{keybox}[Key Implementation Considerations]
$\bullet$ Collecting \blue{high-quality preference data} at scale\\[2pt]
$\bullet$ Ensuring \green{privacy protection} and consent for sensitive musical preference data\\[2pt]
$\bullet$ Addressing \orange{bias} in rater demographics and cultural representation\\[2pt]
$\bullet$ Designing \purple{intuitive user interfaces} that enable effective preference specification\\[2pt]
$\bullet$ Maintaining \blue{system explainability} and quality assurance
\end{keybox}

Storage and retrieval systems for large-scale preference datasets present unique challenges given the multimodal nature of musical data. Unlike text preferences, musical preferences often require storing both symbolic representations such as \dataset{MIDI} \citep{melechovsky2024midicaps} or \dataset{audio} \citep{roy2025jamendomaxcaps}, creating substantial storage requirements. Efficient indexing schemes must enable rapid retrieval of similar musical content for preference model training while supporting diverse query types including harmonic similarity, rhythmic patterns, and stylistic attributes.

Training infrastructure requirements vary significantly across different preference alignment approaches. DiffRhythm+'s integration of \method{DPO} into diffusion architectures demands careful memory management during reverse diffusion processes, as preference optimization requires maintaining gradients through the entire denoising chain. This creates memory scaling challenges that exceed those of standard diffusion training, necessitating techniques such as \method{gradient checkpointing} \citep{chen2016training} and \method{mixed-precision computation} \citep{micikevicius2017mixed} to enable training on practical hardware configurations.

Deployment considerations must account for the distinct computational demands of different preference alignment strategies. Training-time methods like MusicRL add computational costs during development but enable efficient deployment with minimal inference overhead. Conversely, inference-time optimization approaches like Text2midi-InferAlign add computational requirements after deployment, requiring sophisticated resource allocation to maintain consistent response times.

Quality assurance and monitoring present ongoing challenges for deployed preference-aligned systems. Unlike traditional generation systems where quality can be assessed through automated metrics, preference-aligned systems require continuous evaluation of alignment with human preferences that may evolve over time. This necessitates development of automated monitoring systems capable of detecting preference drift and triggering retraining procedures when alignment quality degrades below acceptable thresholds.

\section{Challenges and Open Research Questions}


Despite recent advances, fundamental challenges remain. 

\textbf{Scalability} is critical as current approaches struggle with long-form compositions due to attention complexity and hierarchical structure modeling across temporal scales \cite{zixun2021hierarchical}. 

\textbf{Multi-modal alignment} poses challenges for video-music synchronization and cross-cultural media integration requiring real-time adaptation while maintaining musical coherence \cite{kang2024video2music}.

\textbf{Personalization} remains largely unexplored, requiring \method{few-shot preference learning} methods and efficient representation learning for individual aesthetic judgments \cite{huron2008sweet}. Cultural awareness is essential here, as current systems predominantly reflect Western musical traditions, limiting global applicability and risking cultural appropriation.

\textbf{Computational efficiency} challenges stem from inference-time optimization overhead, energy consumption, and latency requirements for interactive applications.

The fundamental challenge of preference representation learning emerges from MusicRL's findings that traditional metrics capture only partial aspects of human musical judgment. This gap indicates the need for more sophisticated preference representation learning that can capture implicit musical understanding, emotional responses, and contextual appropriateness that human evaluators demonstrate.

\textbf{Cross-domain generalization} presents significant challenges as evidenced by the domain-specific nature of current evaluation frameworks. DiffRhythm+'s reliance on \metric{SongEval} and \metric{Audiobox-aesthetic} metrics highlights the limitation of genre-specific evaluation approaches, while Text2midi-InferAlign's \metric{CLAP-based optimization} may not generalize effectively across diverse musical styles and cultural contexts. The development of universal preference alignment techniques that maintain effectiveness across musical genres, cultural traditions, and application domains remains an open research question.

\textbf{Dynamic preference adaptation} represents a critical gap in current methodologies. While few approaches demonstrate \method{inference-time optimization} capabilities, the field lacks strategies for adapting to evolving user preferences over time. This challenge encompasses both technical aspects—developing efficient incremental learning algorithms for preference models—and theoretical foundations for understanding how musical preferences change through exposure, expertise development, and cultural evolution.

The \textbf{evaluation paradox} in preference-aligned systems creates circular dependencies where assessment of preference alignment quality itself requires human judgment, potentially introducing the same biases and limitations that preference alignment aims to address.



\section{Vision and Roadmap}

\textbf{Applications:} Preference-aligned systems can enable interactive composition tools, adaptive film scoring, game audio, therapeutic music generation \cite{agres2021music}, and personalized music services.

\noindent\textbf{Technical Priorities:} (1) Creating open, large-scale preference datasets spanning diverse cultures and personalization dimensions; (2) developing unified inference-time frameworks for multi-objective optimization with reduced overhead; (3) building cross-cultural evaluation frameworks and culturally-competent models through collaboration with ethnomusicologists; (4) enabling real-time adaptive systems for interactive human-AI co-creation.

Success requires interdisciplinary collaboration combining machine learning, music theory, cognitive science, and ethics to serve diverse creative needs.

\section{Conclusion}

This paper advocates for systematic preference alignment in music generation, addressing the gap between computational optimization and human appreciation. Through examining \method{large-scale preference learning}, \method{inference-time optimization}, and \method{multi-preference diffusion}, we identified promising directions and challenges.

\begin{highlightbox}[Key Takeaways]
Preference alignment improves \textbf{quality and relevance} while preserving \textbf{diversity}. Realizing full potential requires addressing scalability, culturally-aware frameworks, and multi-modal capabilities.
\end{highlightbox}

We call for interdisciplinary research combining machine learning, music theory, and human-computer interaction to create music AI serving human creative needs. The future lies in meaningful alignment with human musical preference and cultural expression.

\section*{Acknowledgements}
This work has received support from SUTD’s Kickstart Initiative under grant number SKI 2021 04 06 and MOE under grant number MOE-T2EP20124-0014.










\bibliography{aaai2026}

\end{document}